\documentstyle[aps,prd,twocolumn,epsf]{revtex}
\begin{document}
\draft
\title{Dark matter phase space densities}
\author{Jes Madsen}
\address{Institute of Physics and Astronomy, University of Aarhus, 
DK-8000 \AA rhus C, Denmark}
\date{June 6, 2000; scheduled to appear in Physical Review D 15July01}
\maketitle

\begin{abstract}
The low velocity part of a kinetic equilibrium dark matter 
distribution has higher phase
space density and is more easily incorporated in formation of a low mass
galaxy than the high velocity part. 
For relativistically decoupling fermions (bosons), 
this explains one (two) orders of magnitude of the observed trend, that
phase space densities in dark matter halo cores are highest 
in the smallest systems,
and loosens constraints on particle masses significantly. 
For non-relativistic decoupling and/or finite chemical potentials
even larger effects may occur.
It is therefore premature to dismiss dissipationless
particle distributions as 
dark matter on the basis of phase space arguments. 
\end{abstract}


It has recently become clear, that the otherwise successful cold dark
matter model (CDM) for cosmic structure formation has several severe
problems, such as predictions of cusps in the central density profiles
of galaxies, too many low-mass subclumbs within dark matter halos, and
a lack of angular momentum in galaxy disks compared to observations. In
one way or the other these problems are all related to the fact
that CDM has no initial velocity spread, and therefore infinite
initial density in phase space. Selfinteractions among the dark matter
particles has been suggested as a possible solution\cite{speste00}. 
Another suggested
solution has been the reintroduction of warm dark matter (WDM) consisting of
particles with a moderate primordial velocity spread\cite{pagpri82}.
Contrary to CDM thermal WDM particles lead to finite
core density, but apparently require a high mass and therefore extremely
early decoupling from the primordial plasma to account for the
observed core phase space density of $10^{-4}M_\odot/{\rm pc^3(km/s)^{-3}}$ in
dwarf spheroidal galaxies.
There may also be difficulties explaining
the decrease in phase space density by a factor 10--100 for dwarf spirals 
and low surface brightness
galaxies, and a further factor of 10--100 for normal 
spirals\cite{sel00,dalhog00}. A single
particle mass for WDM would appear to lead to a definite prediction for
the central halo phase space density because of conservation of
fine-grained phase space density (Liouvilles theorem).

However, whereas such arguments are correct in terms of the average phase
space density, they do not take into account, that while originally
almost uniform in real space, the fine-grained density is an
(exponentially) varying function of position in momentum space. Low
momentum particles are in denser parts of phase space than high momentum
particles, and depending on the actual distribution function (fermion or
boson, zero or nonzero chemical potential, relativistic or
nonrelativistic decoupling), the densest part of the distribution may
have phase space density significantly above average. Furthermore,
halo formation should typically include particles from the low momentum
end first. A system with low gravitational potential like a dwarf
spheroidal will effectively probe only the densest part of the phase
space distribution, whereas a large spiral galaxy probably contains
something close to the average phase density (possibly diluted by
mergers etc).

Other effects (baryons, mergers, phase space dilution during
gravitational collapse, particle selfinteractions etc) can further 
enhance the diversity in observed phase space
densities in galaxy cores, and as demonstrated in \cite{dalhog00} a
merging hierarchy where larger systems are gradually formed by merging
of smaller units can explain many of the features observed.
But as demonstrated in the following, even in the absence of mergers,
a significant spread in primordial phase space densities in dark matter
cores is predicted, with core densities decreasing with increasing escape
velocity of the system in question.
The magnitude of the effect ranges from one order of
magnitude for relativistically decoupling, non-degenerate fermions, to
several orders of magnitude for non-relativistically decoupling bosons.
Ad hoc additions to these simplest distribution functions would
allow a further range. Thus, there may still be room for a single
dissipationless elementary particle explanation of dark matter in dwarf
spheroidal as well as dwarf spiral galaxies\cite{exp}. Furthermore, the natural
selection of the highest phase density particles in the smallest systems
leads to a significant reduction in the minimum mass for the particle
responsible, loosening the rather strong requirements on the epoch of
dark matter decoupling in the early Universe.

An isotropic gas of particles in kinetic equilibrium has a spatial
number density 
\begin{equation}
n=(g/h^3)\int 4\pi f(p) p^2dp, 
\end{equation} 
where $h$ is Planck's
constant, $g$ is the number of helicity states, and
\begin{equation}
f(p)=\left\{ \exp \left[(E-\mu )/kT \right] \pm 1 \right\} ^{-1}
\label{disteq}
\end{equation} 
is a Fermi-Dirac ($+$) or Bose-Einstein ($-$) distribution with chemical
potential $\mu$. Energy $E$ is related to momentum $p$ and particle mass
$m$ via $E^2=(pc)^2+m^2c^4$, where $c$ is the speed of light.

A particle species that decouples from the remaining plasma in the early
Universe at temperature $T_D$ redshifts its momenta in proportion to
the expansion of the Universe, $p=p_D R_D/R$, where $R(t)$ is the cosmic
scale factor, and its number density (particle number is conserved)
evolves like $n\propto R^{-3}$. From this follows that the distribution
function $f$ at a time after decoupling is related to the distribution
at decoupling $f_D$ from Eq.~(\ref{disteq}) like
\begin{equation}
f(p)=f_D(pR/R_D).
\label{fconseq}
\end{equation}
$f(p)$ even keeps an equilibrium shape after decoupling in two regimes
\cite{koltur},
namely ultrarelativistic decoupling ($E\approx pc$, $T=T_DR_D/R$,
$\mu=\mu_D R_D/R$), with
\begin{equation}
f_R(p)=\left\{ \exp \left[(pc-\mu )/kT \right] \pm 1 \right\} ^{-1},
\end{equation} 
and nonrelativistic decoupling ($E-\mu \approx p^2/2m -\mu_{\rm kin}$, 
$\mu_{\rm kin}\equiv \mu-mc^2 =\mu_{\rm kin,D}(R_D/R)^2$, $T=T_D(R_D/R)^2$), so
\begin{equation}
f_N(p)=\left\{ \exp \left[(p^2/(2m)-\mu_{\rm kin} )/kT \right] \pm 1 \right\}
^{-1}.
\end{equation} 

The distribution of fine-grained phase space density is conserved in
time (Liouvilles theorem). This fact is expressed by
Eq.~(\ref{fconseq}), and it means that the phase space distribution at
decoupling can be directly related to measurements of dark matter phase
densities today. Applied to conservation of the maximum phase space
density this was the basis for the Tremaine-Gunn limit on dark matter
fermion masses\cite{tregun79}, later generalized to bosons (where no maximum
exists) by means of the average phase space density\cite{mad90}.

Recently Hogan and Dalcanton \cite{hogdal00} reconsidered the issue of
the primordial average phase space density compared with observations of
dark matter phase space densities in halo cores.
From the distribution function they calculated mass density $\rho=mn$,
and pressure $P=(g/h^3)\int p^2/(3E)f d^3p\approx (g/h^3)\int
p^4/(3mc^2)4\pi dp =mn<v^2>/(3c^2)$, performing the calculation in the
nonrelativistic regime, where $E\approx mc^2$ and $p=mv$. This leads to
the following expression for the phase space density, $Q$, defined by
\begin{equation}
Q\equiv {{\rho}\over{<v^2>^{3/2}}} = 
{{m^4 g}\over{h^3}} 4\pi {{[\int f(p)p^2dp]^{5/2}}
\over {[\int f(p)p^4dp]^{3/2}}} .
\end{equation}

Hogan and Dalcanton \cite{hogdal00} uses this expression in the form
(units with $\hbar=c=1$)
\begin{equation}
Q_X=q_X g_Xm_X^4
\end{equation}
for particle type $X$, where $q_X=0.0019625$ for a relativistically
decoupling, $\mu=0$ fermion, and $q_X=0.036335$ for a relativistically
decoupling, $T=0$, $\mu \gg m$ degenerate fermion, where the $q$-values
come from taking the complete integrals over the distribution function.

For the remainder of this investigation we return to dimensional units and keep
factors of $\hbar$ and $c$. In these units 
\begin{equation}
q_X\equiv Q_Xh^3m^{-4}g^{-1} = 4\pi {{[\int f(p)p^2dp]^{5/2}}\over
{[\int f(p)p^4dp]^{3/2}}} .
\label{qxeq}
\end{equation}
For a degenerate fermion with relativistic decoupling (limit $(m-\mu
)/kT_D \rightarrow -\infty$) this gives
$q_{RFdeg}=4\pi 5^{3/2}/3^{5/2}=9.0128415$, so $\bar f=q/9.0128415$ 
is in general a
measure of the average occupation number in a phase space distribution
(in units of $g/h^3$). For a zero chemical potential fermion the
corresponding average $q$-values for relativistic and non-relativistic
decoupling (in the latter case zero chemical potential means $\mu_{\rm
kin} =0$) are $q_{RF0}=0.4868039$ and $q_{NF0}=1.9223$, whereas the
similar numbers for bosons are $q_{RB0}=0.9071055$ and $q_{NB0}=21.521$.
Notice the larger values for bosons, that express the fact that bosons
have a higher fraction of low momentum, high phase space density
particles than fermions.

So far only average characteristics of the distributions have been
discussed, but it turns out to be quite interesting to study the whole
distribution of phase densities. Figure 1 shows these distributions in a
plot of $q(p)/q_X$ as a function of the fraction of particles with
momentum less than $p$ (calculated at decoupling, but the distribution
is conserved in time by Liouvilles theorem). Here $q_X$ is given in
Eq.~(\ref{qxeq}) integrating from 0 to $\infty$, whereas $q(p)$ is
defined by the same equation, but integrating only from 0 to $p$\cite{thl}. 
The ratio therefore illustrates the amplification of phase space density
relative to the average for a given dark matter distribution function if
only the densest parts of phase space are utilized, for instance in
formation of a galaxy halo.

Notice that relativistically decoupling fermions show an order of
magnitude amplification, whereas two orders of magnitude can be gained
for relativistically decoupling bosons and several orders of 
magnitude for nonrelativistically decoupling bosons.

Another way of illustrating the amplification effect is shown in Figure
2, where amplification is plotted as a function of the dimensionless
momentum $x\equiv p_Dc/kT_D$, which is the natural integration variable
in the calculations. For particles that are nonrelativistic at the epoch
of galaxy formation, $R_g$, $x$ is related to particle speed $v$ at
that time by
\begin{equation}
x={v \over c}{{mc^2}\over{kT_D}}{{R_g}\over{R_D}} .
\end{equation}

\begin{figure}
\epsfxsize=8.6truecm
\epsfbox{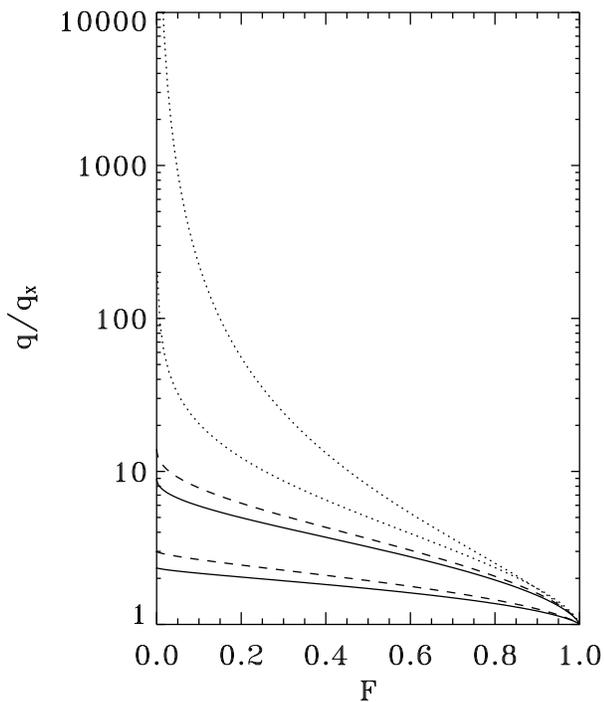}
\caption[]{
Phase space density of dark matter particles in units of the mean
density as a function of fraction of particles, $F$. Upper (lower) 
solid curves are
fermions with $(m-\mu_D) /T_D=0$ and $m/T_D\rightarrow 0$ ($\infty$).
Lower (upper) dotted curves are for bosons in the same limits. 
Dashed curves are for fermions and bosons alike in the limits 
$(m-\mu_D)/T_D\rightarrow \infty$ for $m/T_D\rightarrow 0$ (upper) and
$m/T_D\rightarrow\infty$ (lower).
Fully degenerate fermions ($(m-\mu_D)/T_D\rightarrow -\infty$) have no
amplification factor, i.e. $q/q_X\equiv 1$.
}
\label{fig:1}
\end{figure}

\begin{figure}
\epsfxsize=8.6truecm
\epsfbox{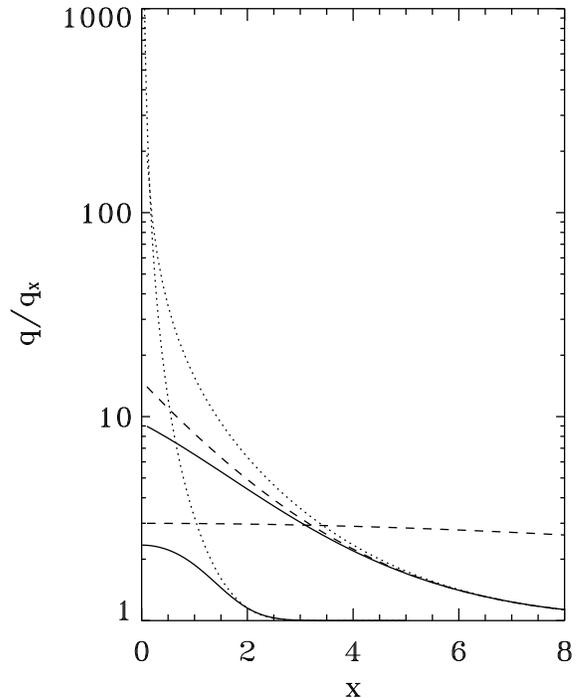}
\caption[]{
As Figure 1, but as function of the dimensionless momentum, $x$.
}
\label{fig:2}
\end{figure}

The occurrence of a phase space amplification factor at low momenta is
a natural consequence of Eq.~(\ref{disteq}). The fine-grained occupation
number (and therefore also the coarse-grained phase space occupation)
has a maximum at $p=0$ equal to $f_{\rm max}=\left\{ \exp \left[(mc^2-\mu
)/kT \right] \pm 1 \right\} ^{-1}$, which is 1 for degenerate fermions
($(mc^2-\mu_D)/kT_D\rightarrow -\infty$), 0.5 for fermions
decoupling when $(mc^2-\mu_D)/kT_D =0$, diverges for bosons in the same
limit, and equals $\exp\left[ (\mu_D-mc^2)/kT_D \right]$ for fermions and
bosons in the limit $(mc^2-\mu_D)/kT_D\rightarrow \infty$.
Quantitatively, the amplification factor for a relativistically
decoupling fermion behaves like $9.26(1-5x/16)$ to first order in $x$
(Fig.~\ref{fig:2}), or $9.26(1-0.691F_F^{1/3}$) expressed in terms of
the fraction of fermions, $F_F$ (Fig.~\ref{fig:1}). The similar limits
for bosons are $19.59 x^{-1}(1-7x/30)$, or $8.93F_B^{-1/2}-4.57$ (notice
that these factors diverge for small $x$ or $F_B$).

Using entropy conservation in the cosmic expansion $R_g$ and $R_D$ 
entering the equation for $x$ are
related by $g_{*g}T_{\gamma g}^3R_g^3=g_{*D}T_{\gamma D}^3R_D^3$,
where $g_{*}$ counts the total number of effective particle degrees of
freedom at the given epoch. 
Today and at galaxy formation, $g_{*g}=43/11$.
Introducing the redshift of galaxy formation
$z_g$ via $T_{\gamma g}=(1+z_g)T_{\gamma 0}$, where the present photon
temperature is $T_{\gamma 0}=2.726{\rm K}$, $x$ can be expressed as
\begin{equation}
x=0.0223 \left( {{mc^2}\over{1{\rm eV}}}\right)
\left( {{10}\over{1+z_g}}\right) 
\left( {v\over{10 {\rm km/s}}}\right)
g_{*D}^{-1/3} .
\end{equation}

The value of $g_{*D}$ depends on the epoch of
decoupling. For standard neutrino decoupling at $kT_D\approx 1{\rm MeV}$,
$g_{*D}=43/4$. But
much higher values of $g_{*D}$ are possible for earlier decoupling,
with $g_{*D}\approx 50$ above the quark-hadron phase transition
temperature, $kT\approx 100{\rm MeV}$, $g_{*D}\approx 100$ above
$kT\approx 200 {\rm GeV}$, and even higher values possible earlier on.

The value of $g_{*D}$ not only determines the $x$-$v$ relation, but
also crucially impacts on the total mass density contribution of the
particle in question, and thereby determines its potential as a dark
matter candidate.
For the case of relativistic decoupling ($(m-\mu_D)/T_D=0$,
$m/T_D\rightarrow 0$) fermions contribute to the cosmic density like
\begin{equation}
\Omega_Xh^2= 0.0572 \left( {{mc^2}\over {1 {\rm eV}}}\right) 
g/g_{*D} ,
\end{equation}
where $h$ is now the Hubble-parameter in units of 100 km~s$^{-1}$~Mpc$^{-1}$,
and a similar expression (multiplied by $4/3$) applies for bosons.

It is interesting to note, that an explanation of the highest phase
space densities measured, those in dwarf spheroidals of order $10^{-4}
M_\odot {\rm pc^{-3}(km/s)^{-3}}$, for a $g=1$ boson requires a particle
mass of only 224~eV (309~eV) if the densest 1\% (10\%) of the bosons are used,
increasing to 681~eV for the average occupation number (masses are
reduced by a factor $2^{-1/4}$ if $g=2$). Such masses are not in
conflict with estimates of cosmic density for reasonably high values of
$g_{*D}$, and $x$ can indeed be low enough for reasonable values of $z_g$
to select only the densest part of phase space for dwarf spheroidals
with typical velocity dispersions below 10~km/s. 

For fermions with $g=2$ the high-phase space density selection expected for
formation of dwarf spheroidals reduces warm dark matter particle mass
limits from 669~eV to 383~eV, again loosening constraints on $g_{*D}$
\cite{scale}.

Even stronger effects may occur in a non-relativistic decoupling regime (upper
dotted curve in Fig.~1),  or one might consider adding extra features to
the dark matter distribution functions, such as a small amount of
Bose-Einstein condensation in the zero momentum, infinite phase density
part of a boson distribution. 

Such ``fine-tuning'' may further loosen constraints on dissipationless dark
matter, but even without it, part of the observed trend for the
core phase space density to decrease with increasing gravitational
potential when going from dwarf spheroidal to dwarf spiral galaxies
is naturally explained by the
selection of low-momentum dark matter particles described here\cite{exp}.
At the same time constraints on WDM particle masses and decoupling
epochs are significantly reduced.
These considerations should be taken into account and tested in 
detailed numerical
simulations of dark matter halo formation, which are needed to settle the
question of whether dissipationless particles may after all account for
the dark matter in galaxies.

This work was supported in part by the Theoretical Astrophysics Center
under the Danish National Research Foundation.

\begin{references}
\bibitem{exp} Formation of larger galaxies and clusters clearly have to
involve significant phase space dilution, probably because of
mergers\cite{dalhog00}.
\end{references}
\end{document}